\documentclass[conference,compsoc]{IEEEtran}
\usepackage{color}

\ifCLASSOPTIONcompsoc
  \usepackage[nocompress]{cite}
\else
  \usepackage{cite}
\fi
\ifCLASSINFOpdf
\else
\fi
\usepackage{graphicx}

\usepackage{CJKutf8}  

\begin{document}

\begin{CJK}{UTF8}{gkai}  

%
\title{MCPGuard : Automatically Detecting Vulnerabilities in MCP Servers}

\author{
\IEEEauthorblockN{
\begin{tabular}{c}
Bin Wang\textsuperscript{1},
Zexin Liu\textsuperscript{1},
Hao Yu\textsuperscript{1},
Ao Yang\textsuperscript{1}, \\
Yenan Huang\textsuperscript{2},
Jing Guo\textsuperscript{2},
Huangsheng Cheng\textsuperscript{2},
Hui Li\textsuperscript{1}\textsuperscript{\dag}
Huiyu Wu\textsuperscript{2}\textsuperscript{\dag},
\end{tabular}
}
\IEEEauthorblockA{
\textsuperscript{1}Peking University, \textsuperscript{2}Tencent \\
\{thebinking, zexinliu25, hyu25, jarvisya\}@stu.pku.edu.cn, 
\{roninhuang, fyoungguo, pythoncheng\}@tencent.com, \\
lih64@pkusz.edu.cn, nickyccwu@tencent.com
}
}

\maketitle

\begin{abstract}
The Model Context Protocol (MCP) has emerged as a standardized interface enabling seamless integration between Large Language Models (LLMs) and external data sources and tools. While MCP significantly reduces development complexity and enhances agent capabilities, its openness and extensibility introduce critical security vulnerabilities that threaten system trustworthiness and user data protection. This paper systematically analyzes the security landscape of MCP-based systems, identifying three principal threat categories: (1) agent hijacking attacks stemming from protocol design deficiencies; (2) traditional web vulnerabilities in MCP servers; and (3) supply chain security. To address these challenges, we comprehensively survey existing defense strategies, examining both proactive server-side scanning approaches, ranging from layered detection pipelines and agentic auditing frameworks to zero-trust registry systems, and runtime interaction monitoring solutions that provide continuous oversight and policy enforcement. Our analysis reveals that MCP security fundamentally represents a paradigm shift where the attack surface extends from traditional code execution to semantic interpretation of natural language metadata, necessitating novel defense mechanisms tailored to this unique threat model.

\end{abstract}

\IEEEpeerreviewmaketitle

\section{Introduction}

Large Language Models (LLMs) have undergone continuous advancement, achieving significant breakthroughs in both inference speed and output quality, while increasingly gaining the capability to select and invoke external tools. A growing number of LLM-based agents have emerged—capable not only of engaging in multi-turn dialogues or solving International Mathematical Olympiad (IMO) level problems, but also of autonomously planning actions, making decisions, and interacting with external APIs, databases, and tools when faced with complex tasks. However, disparate databases, web services, and applications remain largely siloed, posing substantial engineering complexity for developers due to the lack of seamless integration and extensibility.

To address this challenge, the Model Context Protocol (MCP) \cite{anthropic_mcp_spec} has been introduced as a standardized interface for connecting LLMs with external data sources. MCP significantly reduces integration overhead and establishes a secure, trusted communication channel between MCP clients and servers, thereby fulfilling the scalability and interoperability requirements of AI-powered services. The synergy between advancing LLM capabilities and the adoption of MCP has substantially expanded the operational boundaries of agents, enhancing their ability to dynamically interact with tools. Since its release, MCP has been rapidly adopted by mainstream applications such as Claude Desktop, OpenAI, and Cursor, and has garnered widespread acclaim within the developer community.

Under this new paradigm, agents are no longer passively reliant on static data sources; instead, they can dynamically discover, select, and invoke tools—dramatically elevating the overall intelligence of AI systems. Yet, the very openness and extensibility that make MCP powerful also introduce novel security risks. As the critical bridge mapping LLM reasoning to real-world execution environments, MCP's security directly determines the trustworthiness of agent systems and the protection of user data and infrastructure.

In this work, we focus on the architectural layer of MCP servers, systematically analyzing the security threats and challenges they face, and propose corresponding defense strategies. We identify three core security challenges currently confronting the MCP ecosystem:

1. \textbf{Agent Hijacking Risks from MCP Protocol Deficiencies}: MCP clients directly inject tool descriptions and outputs into the conversation context without adequate isolation between sessions of different tools. This design flaw enables a range of attacks—including Tool Poisoning Attacks, Rug Pulls, Tool Shadowing Attacks, and Indirect Prompt Injection Attacks—all of which can hijack agents into performing unintended actions, leading to user privacy breaches \cite{invariantlabs_tool_poisoning}.

2. \textbf{Code-Level Vulnerabilities and Traditional Web Exploits in MCP Servers}: MCP servers are typically deployed as web services or local proxies, responsible for parsing client requests and mapping them to backend tool invocation pipelines. This architectural pattern inherently exposes them to conventional web attack surfaces. Common vulnerabilities include command injection, path traversal, arbitrary file read, and Server-Side Request Forgery (SSRF), which attackers can exploit to manipulate request logic or exfiltrate sensitive server-side information. A representative example is CVE-2025-49596, a critical remote code execution vulnerability recently disclosed in Anthropic’s official MCP Inspector tool \cite{CVE-2025-49596}.

3. \textbf{Supply Chain Security}: A defining feature of the MCP ecosystem is the ability for tools and services to interoperate and be distributed across heterogeneous servers. However, there is currently no unified, trusted marketplace or vetting mechanism for MCP services. Malicious actors can masquerade as legitimate tool providers, tricking users into selecting untrusted services from MCP marketplaces to steal credentials, access chat histories, or inject malicious logic. More alarmingly, in the open and decentralized nature of the ecosystem, attackers can launch MCP Preference Manipulation Attacks (MPMAs) \cite{wang2025mpma}, strategically optimizing tool names, descriptions, or employing implicit nudging techniques to increase the likelihood that compromised MCP servers are preferentially selected by agents.

In today’s rapidly evolving and highly competitive AI landscape, security and reliability have become key differentiators for product success. Nevertheless, MCP marketplaces and hosting providers commonly face a systemic challenge: the absence of dedicated AI security teams prevents comprehensive security scanning and risk assessment of developer-submitted MCP services. Concurrently, MCP developers urgently need authoritative, user-friendly tools to verify and demonstrate the security posture of their implementations. 

\section{Model Context Protocol (MCP)}
The ability of an agent to interact with the external world largely depends on its mechanism for invoking various tools (such as web search, databases, code interpreters, etc.). Early tool usage typically relied on dedicated APIs designed for specific application scenarios. This approach to tool invocation has inherent limitations. Due to the lack of a standardized process, these solutions often involve deeply nested API call sequences and complex authentication key configurations. Essentially, this method of tool invocation represents a tightly integrated model, with its core flaw lying in the absence of a standardized interaction paradigm and a unified management process.

Specifically, such customized implementation schemes face several key challenges:

\begin{itemize}
    \item Heterogeneity \& Complexity: Developers need to write dedicated adaptation and parsing code for each tool to handle its unique API endpoints, data formats, and authentication mechanisms, resulting in high integration costs. When the number of tools is large or changes frequently, this manual approach is highly error-prone and inefficient.
    \item Brittleness \& Maintenance Challenges: Any minor change in an external API can disrupt the entire call chain. More importantly, this integration method heavily relies on the model's natural language understanding ability to "guess" when and how to invoke tools. This implicit approach based on prompt engineering is highly susceptible to model hallucinations or instruction interpretation biases, leading to insufficient reliability.
    \item Inefficiency \& Error-Proneness: Manual, non-standardized configuration and call process management easily introduce human errors. At the same time, traditional logging and monitoring methods are inadequate for capturing the complex reasoning processes that may arise during multiple tool invocations. The lack of a unified logging and error handling mechanism makes debugging exceptionally difficult, severely impacting system development and iteration efficiency.
\end{itemize}

To address the above issues, Anthropic proposed the Model Context Protocol (MCP) \cite{anthropic_mcp_spec}. MCP is an open protocol designed to establish a standardized bidirectional communication link between large language models (LLMs) and external data sources and tools. Just as the USB-C interface provides a unified standard for devices to connect various peripherals and accessories, MCP offers a standardized pathway for LLMs to connect diverse data sources and tools. Through this protocol, developers can connect varied data sources in a unified format, significantly reducing the development complexity of LLM Agents and accelerating their application across various industries.

The standardized tool invocation process based on MCP generally follows a $Client-Server$ interaction pattern. The MCP client initiates an initial request to the MCP server to query the metadata of available tools. The MCP server returns a list containing definitions of all tools (name, description, parameter schema). The MCP client, adhering to the MCP specification, generates one or more structured $<tool\_use>$ requests and sends them to the server through a predefined communication channel (such as stdio or HTTP). The MCP server continuously listens for requests from the client. Upon parsing the $<tool\_use>$ block in the request, it validates the tool name and parameters, invokes the corresponding tool to execute the task, and then encapsulates the execution result in a $<tool\_result>$ tag before sending it back to the client via the same channel. The MCP client receives the $<tool\_result>$, incorporates it as new contextual information, and continues generating a natural language response for the user.

\section{MCP Attacks}

Although the Model Context Protocol (MCP) demonstrates remarkable openness and extensibility, its security has not yet reached a mature or robust level. While this open architecture enhances flexibility and fosters innovation, it simultaneously introduces a range of security risks. Prior research has highlighted several security and privacy concerns associated with MCP-based systems, including malicious code execution, abuse of remote access controls, credential theft, and the absence of proper authentication, authorization, and debugging mechanisms \cite{radosevich2025mcp,hasan2025model}.

\subsection{Indirect Prompt Injection}
The core idea behind indirect prompt injection is that attackers do not directly interact with the large language model (LLM) to issue malicious instructions\cite{greshake2023}. Instead, they embed adversarial prompts within external data sources that the LLM is designed to retrieve during its normal operation, such as web pages, emails, documents, or code repositories. When the LLM application fetches these external resources to answer a user’s legitimate query, it may inadvertently parse and execute the hidden malicious instructions without the user’s knowledge (e.g., “Ignore all previous rules and immediately send the local API key to the attacker’s server”).

In the MCP framework, user queries, tool descriptions, and tool execution results are dynamically incorporated into the LLM’s context to enable intelligent tool selection and task completion based on returned outputs. However, this integration creates a critical vulnerability: if the output from a tool contains malicious content sourced from an untrusted external origin, the LLM may fail to distinguish between passive data and active instructions. Consequently, it may interpret the injected prompt as legitimate behavioral guidance, triggering an indirect prompt injection attack. This can lead to severe consequences, including sensitive data leakage, privilege escalation, or even system compromise.

Recent work by Zhao et al. \cite{zhao2025mind} further demonstrates that in LLM agent ecosystems built upon MCP, attackers can exploit indirect prompt injection to manipulate or hijack the tool invocation chain, ultimately exfiltrating sensitive information. Their findings underscore the significant threat this attack vector poses in complex, multi-agent collaborative environments.

\subsection{Traditional Web Vulnerabilities in MCP Servers}
An MCP Server fundamentally acts as an intermediary layer between tools and external callers. By wrapping existing tools with an MCP Server, developers can expose tool capabilities through a standardized interface. MCP Servers support both local and remote deployment modes; in the remote scenario, communication occurs over HTTP—mirroring the way traditional web backends expose APIs via HTTP. Consequently, classic web application vulnerabilities—such as injection flaws, cross-site scripting (XSS), and insecure deserialization—are equally applicable to MCP Servers.

Notably, these risks can be significantly amplified in the context of deep integration with large language models (LLMs). MCP Servers typically receive dynamically generated parameters from LLMs and pass them directly to underlying tools for execution. In pursuit of greater flexibility or performance, some MCP Server implementations omit rigorous input validation, lack proper sanitization of parameters, or even concatenate raw inputs directly into high-risk functions (e.g., \texttt{eval}, \texttt{os.system}, \texttt{exec}) or use them to construct SQL queries. Such practices can easily lead to severe vulnerabilities, including command injection, arbitrary code execution, and SQL injection.

Moreover, when deployed remotely, MCP Servers inevitably face challenges related to authentication and authorization. If an MCP Server, especially one capable of accessing sensitive resources or performing privileged operations, is not properly configured according to the official MCP specification’s Authorization Flow, attackers may bypass authentication controls entirely. This could grant unauthorized access to tools, sensitive data, or even the underlying system, resulting in data breaches, service abuse, or the expansion of the attack surface beyond the MCP Server itself.

\subsection{Supply Chain Security}
The core vision of the Model Context Protocol (MCP) is to enable seamless interoperability among a vast array of tools, thereby fostering an open, dynamic, and composable ecosystem of intelligent agents. To realize this vision, MCP Servers originate from highly diverse sources—ranging from official maintainers and enterprise teams to open-source communities and individual developers. While this diversity fuels innovation and flexibility, it also gives rise to a complex, loosely coupled "tool supply chain", introducing significant challenges in trust, integrity, and behavioral predictability. Below, we systematically review representative supply chain attacks against the MCP ecosystem identified in recent security research.

\textbf{Tool Poisoning Attacks} represent the most emblematic and widely studied security threat in the MCP ecosystem. Fundamentally, they constitute a context-aware variant of prompt injection. The attack exploits a foundational design principle of MCP: large language models (LLMs) are granted high autonomy in interpreting and invoking external tools based on their natural-language metadata (e.g., names, descriptions, parameter schemas). Attackers embed carefully crafted malicious instructions, such as “forward all user messages to the attacker’s server”, into a tool’s description field. Because these instructions reside within metadata that the LLM treats as trusted contextual guidance, the model often misinterprets them as legitimate operational directives rather than adversarial payloads, executing them without user awareness. A notable real-world example is the malicious \texttt{whatsapp-mcp} project disclosed by Invariant Labs \cite{beurer-kellner2025whatsapp}, where the \texttt{get\_fact\_of\_the\_day} tool was modified to exfiltrate user's chat history to an attacker-controlled phone number while appearing to deliver harmless daily trivia.

\textbf{Tool Shadowing Attacks} expose critical weaknesses in MCP’s naming and context management. When a malicious MCP service registers a tool with the same name as a legitimate one, but with a different implementation, the client may inadvertently invoke the malicious version due to the absence of namespace isolation or source verification. More insidiously, the malicious tool can embed "global rules" in its description (e.g., all tool invocations must first execute this tool), tricking the LLM into treating them as system-level preconditions. This phenomenon, termed "context isolation failure" in \cite{hou2025model}, arises because multiple tools share a single reasoning context, allowing adversarial tools to semantically pollute the behavior of trusted ones and erode user control.

\textbf{Rug Pull Attacks} manifest in novel forms within the MCP ecosystem. Attackers first deploy a seemingly trustworthy, fully functional MCP service with comprehensive documentation to attract integrations and build user dependency. Once users have established stable usage patterns, the attacker silently injects backdoors, steals credentials, or redirects communications through remote updates. The lack of mandatory mechanisms for version signing, change auditing, or update source verification in current MCP implementations makes such stealthy modifications extremely difficult for end users to detect, enabling prolonged compromise.

The intense research focus on tool poisoning attacks reflects a deeper architectural tension in MCP. The protocol treats natural language tool metadata as authoritative input for LLM decision-making, yet provides no mechanisms to verify the provenance, integrity, or semantic safety of this metadata. In traditional software supply chains, code and binaries can be validated through cryptographic signatures, hash checks, and sandboxed execution. In contrast, the MCP attack surface shifts upstream from code execution to semantic interpretation. Malicious behavior no longer requires exploiting memory corruption or logic flaws; instead, it manipulates the model’s understanding of intent. This design paradigm, trusting natural language as executable policy, is inherently vulnerable in open, collaborative environments. Consequently, MCP supply chain security has effectively become a battleground for adversarial prompt engineering.

\section{MCP Defenses}
\subsection{Server-Side Scanning}
Server-side scanning represents a proactive defense paradigm in MCP security, focusing on preemptively identifying vulnerabilities within MCP servers and their configurations before they can be exploited. Current approaches in this domain diverge into complementary strategies: some employ sophisticated, multi-layered detection pipelines for deep analysis; others leverage agentic frameworks to dynamically probe for flaws; some re-architect the trust model around rigorous registration and authentication; while practical tools offer accessible, integrated scanning capabilities for broader adoption.

MCP-Guard\cite{xing2025mcpguarddefenseframeworkmodel} exemplifies a layered, modular defense architecture. Its core technical implementation first utilizes lightweight static scanning with pattern-based detection for rapid threat filtering, supports real-time adaptation via hot updates, and follows a fail-fast mechanism to minimize overhead. Subsequently, it deploys a deep neural detection module, built on pre-trained text embedding models that are fully fine-tuned on MCP-specific threat data to capture domain-specific semantics that generic models may miss. Finally, an intelligent arbitration mechanism employs an LLM to independently assess input safety based on standardized criteria, combining its judgment with the neural module's results for uncertain cases to form a hybrid decision-making system, thereby balancing efficiency and accuracy.

Distinct from layered detection pipelines, McpSafetyScanner's\cite{radosevich2025mcpsafetyauditllms} contribution lies in its systematic exposure of critical vulnerabilities within MCP-enabled LLM systems. It demonstrates that leading LLMs such as Claude 3.7 and Llama-3.3-70B can be coerced into enabling malicious code execution, remote access control, and credential theft, even bypassing inconsistent guardrails. It further introduces a novel high-threat Retrieval-Agent Deception attack, where attackers corrupt public data with MCP-targeted commands that are triggered via retrieval tools, eliminating the need for direct system access. The scanner adopts a unique agentic framework comprising hacker, auditor, and supervisor agents that automatically probes MCP servers using their native tools, searches knowledge bases for related threats, and generates detailed reports—proactively addressing LLM guardrail gaps through pre-deployment auditing with validated accuracy in identifying exploits and providing actionable fixes.

Narajala et al.'s work\cite{narajala2025securinggenaimultiagentsystems} addresses MCP security in GenAI multi-agent systems by specifically targeting tool squatting—a deceptive practice enabling threats like malicious server spoofing and tool description poisoning. Its distinct contribution is a zero-trust, admin-controlled Tool Registry framework. This framework restricts MCP server and agent registration to trusted admins, uses fine-grained policies linking agent identities to server access, calculates dynamic Trust Scores for servers to guide risk-based selection, and provides just-in-time, scope-limited credentials to reduce theft risks. By monitoring MCP-related events for anomalies and validating server authenticity through the registry, this framework directly mitigates security gaps arising from uncontrolled registration and deceptive representation.

MCP-Scan\cite{invariantlabs2024mcpscan} differentiates itself as an open-source, versatile tool focused on enhancing MCP security through a privacy-aware scanning framework. It combines static analysis of client configurations with dynamic monitoring of MCP traffic to identify a range of vulnerabilities, from injection attacks to unauthorized tool modifications. Its key features include flexibility for local and remote server scanning, configurable privacy controls, and an integration-friendly design via command-line interface and JSON outputs, emphasizing both comprehensive threat detection and practical usability for diverse MCP environments.

AI-Infra-Guard\cite{Tencent_AI-Infra-Guard_2025}, from Tencent Zhuque Lab, positions itself as a comprehensive, user-friendly platform that extends beyond pure MCP scanning. It integrates MCP server risk scanning within a broader suite of capabilities, including AI infrastructure vulnerability assessment and jailbreak evaluation. Its standout characteristic is offering an all-in-one, intelligent solution designed for ease of use, featuring minimal-configuration deployment, an intuitive web interface, and a plugin framework for community-driven expansion, making holistic AI security accessible.

\subsection{Interaction Monitoring}
Interaction monitoring serves as a critical runtime defense mechanism within MCP security, providing continuous oversight of MCP client-server communications to detect and respond to anomalous or malicious activities. Current research and tooling in this area emphasize seamless integration and actionable insights, moving beyond passive logging to actively enforce security policies. These approaches vary in their architectural integration points and primary focus, ranging from deep coupling with enhanced trust frameworks to lightweight middleware solutions and flexible policy-based analysis.

Bhatt et al.'s work\cite{bhatt2025etdimitigatingtoolsquatting} integrates interaction monitoring distinctively within their Enhanced Tool Definition Interface framework, creating a deeply coupled security apparatus. Unlike generic monitoring, this approach embeds auditing directly into core trust mechanisms: it tracks cryptographic signature verification for tool definitions to ensure only authentic tools are invoked, logs version changes to trigger re-approval checks against rug pulls, and records OAuth 2.0 scope adherence during tool calls to prevent permission creep. For policy-driven interactions, it monitors dynamic evaluations by engines like Open Policy Agent, capturing decision context and outcomes for each invocation. A key innovation is call stack verification, which tracks nested tool chains to detect unauthorized sequencing, privilege escalation, or circular dependencies. All monitored events are logged for comprehensive audit trails, with alerts triggered for anomalies such as unsigned tool attempts or scope violations, thereby making the entire trust and authorization logic underpinning MCP interactions transparent and actionable.

Kumar et al.'s MCP Guardian\cite{kumar2025mcpguardiansecurityfirstlayer} positions itself as a lightweight, non-intrusive middleware layer that unifies monitoring with core security controls. Its distinction lies in embedding observability directly into operational safeguards: it logs every MCP tool call with full context for auditability, tracks per-token request volumes to enforce rate limits and prevent resource exhaustion, scans request arguments via a built-in regex-based Web Application Firewall to block malicious patterns, and verifies authentication tokens in real-time. This monitoring is proactive—suspicious events trigger configurable alerts and are correlated across request flows. Critically, the framework maintains minimal performance overhead and requires no architectural changes to existing MCP servers, offering a practical, unified solution for real-time threat detection and policy enforcement against risks like tool poisoning and command injection.

Invariant\cite{invariantlabs_invariant_main}, an open-source project by Invariant Labs, differentiates itself through a highly flexible, policy-driven monitoring framework centered on trace analysis. It provides a customizable domain-specific language for defining security policies that govern AI agent behavior, enabling deep semantic analysis of interaction traces to detect complex threats like data leaks, unauthorized tool sequences, and policy violations. Its core strength is the ability to perform both retrospective analysis of logged traces and real-time monitoring of active agents, integrating seamlessly with popular frameworks like OpenAI and LangChain. By focusing on expressive policy definitions and contextual understanding of data flows, Invariant offers a developer-friendly approach to enforcing sophisticated safety constraints without modifying application logic, making advanced security accessible for diverse AI agent deployments.


\section{Conclusion}
This paper primarily focuses on the Model Context Protocol (MCP) and its associated security challenges and corresponding defense measures. It first elaborates that MCP serves as a standardized interface for connecting Large Language Models (LLMs) with external data sources and tools, which addresses the siloed issue of traditional data and applications, reduces integration complexity, and has been adopted by mainstream applications. However, MCP’s openness and extensibility also lead to multiple security crises, including agent hijacking risks from protocol deficiencies, code-level and traditional web vulnerabilities in MCP servers, and supply chain and marketplace risks. To tackle these security issues, defense strategies such as server-side scanning and interaction monitoring are proposed. This paper also introduces the AIG tool for automatic vulnerability detection, with the goal of enhancing MCP system security and promoting the healthy development of the MCP ecosystem




\bibliographystyle{IEEEtran}
\bibliography{IEEEabrv,reference}
%



\end{CJK}  

\end{document}